\documentclass[twocolumn,showpacs,preprintnumbers,amsmath,amssymb]{revtex4}
\usepackage{graphicx}% Include figure files
\usepackage{dcolumn}% Align table columns on decimal point
\usepackage{bm}% bold math

\def\gsim{\mathrel{\raise.3ex\hbox{$>$\kern-.75em\lower1ex\hbox{$\sim$}}}}
\def\lsim{\mathrel{\raise.3ex\hbox{$<$\kern-.75em\lower1ex\hbox{$\sim$}}}}

\begin{document}

\preprint{APS/123-QED}

\title{Thermodynamics of the early Universe\\with mirror dark matter}

\author{Paolo Ciarcelluti}
 \email{paolo.ciarcelluti@ulg.ac.be}
 \affiliation{D\'epartement AGO - IFPA, Universit\'e de Li\`ege, B-4000 Li\`ege, Belgium}

\author{Angela Lepidi}
 \email{lepidi@fe.infn.it}
 \affiliation{Dipartimento di Fisica, Universit\`a di Ferrara, I-41000 Ferrara, 
Italy\\
INFN, sezione di Ferrara, I-44100 Ferrara, Italy}

\date{\today}% It is always \today, today,

\begin{abstract}
Mirror matter is a promising self-collisional dark matter candidate.
Here we study the evolution of thermodynamical quantities in the early Universe for temperatures below $\sim 100$ ~MeV in presence of a hidden mirror sector with unbroken parity symmetry and with gravitational interactions only.
This range of temperatures is interesting for primordial nucleosynthesis analyses, therefore we focus on the temporal evolution of number of degrees of freedom in both sectors.
Numerically solving the equations, we obtain the interesting prediction that the effective number of extra-neutrino families raises for decreasing temperatures before and after Big Bang nucleosynthesis; this could help solving the discrepancy in this number computed at nucleosynthesis and cosmic microwave background formation epochs.
\end{abstract}

\maketitle

%---------------------------------------------------------------------------------

\section{Introduction}

In 50's Lee and Yang~\cite{Lee:1956qn} suggested that the existence of a hidden sector of particles and interactions with exactly the same properties as our visible world (except for right-handed, instead of left-handed, weak interactions) could restore the total symmetry of physical laws.
Many years later Foot at al.~\cite{Foot:1991bp} proposed a model with exact parity (mirror) symmetry between the two sectors, that are described by the same Lagrangian and coupling constants, and consequently have the same microphysics.
Since the only interaction linking the two worlds is gravity\footnote{
There could be other interactions, as for example the kinetic mixing between ordinary and mirror photons~\cite{Foot:2000vy}, but they are very low, and are neglected in the present study.
}, mirror baryons constitute a dark matter candidate in a natural way, as they interact with mirror photons, but not with ordinary ones.

The phenomenology of mirror matter was studied in several papers (for an extended list see the bibliography of~Ref.~\cite{Okun:2006eb}); in particular the implications for Big Bang nucleosynthesis~\cite{Berezhiani:2000gw,HPLW2008}, primordial structure formation and cosmic microwave background~\cite{Ciarcelluti:2003wm,Berezhiani:2003wj,Ciarcelluti:2004ij,Ciarcelluti:2004ik,Ciarcelluti:2004ip,HPLW2008}, large scale structure of the Universe~\cite{Ciarcelluti:2003wm,Berezhiani:2003wj,Ciarcelluti:2004ij,Ciarcelluti:2004ip,HPLW2008,Ignatiev:2003js}, microlensing events (MACHOs)~\cite{Blinnikov:1996fm,Foot:1999hm,Mohapatra:1999ih,Berezhiani:2005vv}, interpretation of DAMA experiment~\cite{Foot:2008nw,Bernabei:2008yi} 
are related to this study.

If the mirror (M) sector exists, then the Universe along with the ordinary (O) particles should contain their mirror partners, but their densities are not the same in both sectors.
Indeed, the BBN bound on the effective number of extra neutrino families implies that the M sector has a temperature lower than the O one, as naturally obtained in some inflationary models~\cite{Berezhiani:1995am}.
Then, two sectors have different initial conditions, they do not come into thermal equilibrium at later epoch and they evolve independently, separately conserving their entropies and maintaining approximately constant the ratio among their temperatures.

All the differences with respect to the O world can be described in terms of only two free parameters, defined as:
\begin{equation} \label{mir-param}
  x \equiv \left( s' \over s \right)^{1/3} \;;
~~
  \beta \equiv \Omega'_{\rm b} / \Omega_{\rm b} \;,
\end{equation}
where $T$ ($T'$), $\Omega_{b}$ ($\Omega'_{b}$), and $s$ ($s'$) are respectively the O (M) photon temperature, cosmological baryon density, and entropy density.\footnote{From now on all particles and parameters of the mirror sector will be marked by $'$ to distinguish them from the ones related to the observable or ordinary world.
}
The bounds on the mirror parameters are $ x < 0.7 $ and $ \beta > 1 $, the first one coming from the previously mentioned BBN limit and the second one from the hypothesis that a relevant fraction of dark matter is made of M baryons.
In general, during most of the history of the Universe, we can approximate
\begin{equation} \label{xmir}
  x \equiv \left( s' \over s \right)^{1/3}
    = \left[q'(T') \over q(T) \right]^{1/3}{T' \over T} \approx {T' \over T} \;,
\end{equation}
where now $q(T)$ and $q'(T)$ are respectively the O and M entropic degrees of freedom, and the M photon temperature $T'(T)$ is a function of the O one.
This approximation is only valid if the temperatures of two sectors are not too different, or alternatively if we are away enough from crucial epochs, like $e^+$-$e^-$ annihilation~\cite{Berezhiani:2000gw}.
But we are just investigating the range of temperatures interested by this phenomenon, and in fact one of the aims of this paper is to exactly compute the trends of these thermodynamical quantities when the previous approximation is no longer valid.

The present study is required for different reasons.
First of all, we need a detailed study of thermodynamical evolution of the early Universe in order to obtain a reliable description of the thermal cosmic history in presence of M dark matter. Secondly, an accurate study of the trend of number of degrees of freedom is necessary for any future simulation of nucleosynthesis of primordial elements in both sectors, that may be compared with observations.
In particular, it becomes even more important in view of the recent proposed interpretation of the DAMA/LIBRA experiment in terms of interactions with M heavy elements~\cite{Foot:2008nw,Bernabei:2008yi}. Thirdly, claims for variations of the effective number of extra neutrino families computed at BBN ($\sim 1$ MeV) and cosmic microwave background (CMB) formation ($\lsim 1$ eV) epochs require investigations for possible variations of these numbers as consequences of physics beyond the Standard Model (see the recent Ref.~\cite{Mangano:2006ur} and references therein for previous works).

The plan of the paper is as follows.
In next section we introduce the thermodynamical equilibrium in the early mirror Universe and the equations that govern its evolution.
In section \ref{sec_Num_calc} we present the numerical solutions of these equations and discuss the results.
Finally, our main conclusions are summarized in section \ref{sec_conclusions}.

%---------------------------------------------------------------------------------

\section{Thermodynamical equilibrium}
\label{sec_eq_therm}

We extend the standard theory of thermodynamics in the early Universe in order to take into account the existence of M particles.

We consider the Universe as a thermodynamical system composed of different species (O and M electrons, photons, neutrinos, nucleons, etc.) which, in the early phases, were to a good approximation in thermodynamical equilibrium, established through rapid interactions, in two sectors separately.

We assume, as usual, that the Universe is homogeneous, and the chemical potentials of all particle species A are negligible, i.e. $\mu_A \ll T$~\cite{Kolb:1990vq,paddybook}.
The latter assumption implies the conservation of the total entropy of the Universe $S$.
Therefore we can use the equilibrium Bose-Einstein or Fermi-Dirac distribution functions and calculate the energy density $\rho$, the pressure $p$, and the entropy density $s$ for every particle species in thermal equilibrium:
\begin{eqnarray} \label{energy_density}
  \rho_A(T) = \frac{g_A}{2\pi^2} \int_{m_A}^{\infty}
    \frac{(E^2-m_A^2)^{\frac{1}{2}} \:E^2}
    {\exp \left[\frac{E}{T}\right]\pm 1} \:dE \;,
\end{eqnarray}
\begin{eqnarray} \label{pressure}
  p_A(T) = \frac{g_A}{6\pi^2} \int_{m_A}^{\infty}
    \frac{(E^2-m_A^2)^{\frac{3}{2}} }
    {\exp \left[\frac{E}{T}\right]\pm 1} \:dE \;,
\end{eqnarray}
\begin{eqnarray} \label{entropy_generic}
  s_A(T) = \frac{p_A(T) + \rho_A(T)}{T} \;,
\end{eqnarray}
where $g_A$ is the spin-degeneracy factor of the species $A$, 
the signs + and -- correspond respectively to fermions and bosons, and $E = \sqrt{\mathbf{p}^2 + m^2}$, with $\mathbf{p}$ the momentum.
We can use the usual parametrization of the entropy density:
\begin{eqnarray} \label{entropy}
  s(T) = \frac{2\pi^2}{45}\: q(T)\: T^3 \;,
\end{eqnarray}
where
\begin{eqnarray} \label{q_tot__def}
  q(T) \equiv&& \sum_{bosons} g_b (T) \left(\frac{T_b}{T}\right)^3 \nonumber\\
    &&+\frac{7}{8}\sum_{fermions} g_f (T) \left(\frac{T_f}{T}\right)^3
\end{eqnarray}
is the number of {\sl effective entropic} degrees of freedom (DoF), that group together the DoF for all bosons ($g_b$) and fermions ($g_f$).
An analogous formalism can be used for the total energy density $\rho$, which can be parametrized as:
\begin{eqnarray} \label{energy}
  \rho (T)&=& \frac{\pi^2}{30}\: g(T)\: T^4 \;,
\end{eqnarray}
where
\begin{eqnarray} \label{g_tot__def}
  g(T) \equiv&& \sum_{bosons} g_b (T) \left(\frac{T_b}{T}\right)^4 \nonumber\\
    &&+\frac{7}{8}\sum_{fermions} g_f (T) \left(\frac{T_f}{T}\right)^4
\end{eqnarray}
is the number of {\sl effective energetic} degrees of freedom.

%.................................................................................

\subsection{Equations}
\label{sec_equations}

We calculate the equations which link the O and M sector temperatures and thermodynamical quantities; then we numerically solve these equations.
Once the temperatures are known, it is possible to work out the total exact 
number of DoF in both sectors, which can be, as it is common in the literature, expressed in terms of extra-neutrino number.
We report in \S \ref{sec_Num_calc} the results of these calculations.

The presence of the other sector leads in both sectors to the same effects of having more particles.
As already stated, we do not take in account interactions between the two sectors besides gravity.
This implies that the entropies of two sectors are separately conserved, and the parameter $x$ is constant during the cosmic evolution:
\begin{eqnarray}
  x =\left(\frac{s'}{s}\right)^{1/3}
    =\left(\frac{s' \cdot a^3}{s \cdot a^3}\right)^{1/3}
    =\left(\frac{S'}{S}\right)^{1/3}
    = const. \;,
\end{eqnarray}
where $a$ is the scale factor.

As already stressed, O and M sectors have the same microphysics; therefore we can assume at first approximation that the neutrino decoupling temperature\footnote{
The assumption that the neutrino decoupling is an instantaneous process taking place when photons have the temperature $T=T_{D\nu}$ introduces just a small error ($< 1\%$), that is negligible for the precision required in this work.
} $T_{D\nu}$ is the same in both of them, that is $T_{D\nu}=T_{D\nu'}'$.
But the temperatures in the O sector when the O and M neutrino decouplings take place are different, $T_{D\nu} \ne T_{D\nu'}$, with $T_{D\nu}<T_{D\nu'}$ since $x<1$.
Therefore, at $T \gg T_{D\nu}$ ($T_{D\nu}'$), O (M) neutrinos are in thermal equilibrium with the O (M) plasma and their temperature is $T_{\nu} = T$ ($T_{\nu}' = T'$), while after decoupling it scales as $a^{-1}$.

In each sector, shortly after the $\nu$ decoupling, the $e^+$-$e^-$ annihilate because the temperature becomes smaller than $2 m_e$, which is the threshold for the reaction $\gamma \leftrightarrow e^+e^-$ in both O and M worlds.
Thus electrons and positrons transfer their entropy to the corresponding photons, which become hotter than neutrinos.

This fact will be used together with the entropy conservations (total and in each sector separately) to find the equations that govern the evolution of the M photon temperature $T'$ as a function of the O one $T$.
In fact, O and M neutrino decouplings are key events, together with both $e^+$-$e^-$ annihilation processes, for the thermodynamics in this range of temperatures.
Once we call $T_{D\nu'}$ the O world temperature when the M neutrino decoupling takes place, we can split the early Universe evolution for temperatures $T \lsim 100$ MeV into three phases.

\subsubsection{Phase~ $T > T_{D\nu'}$}

Photons, electrons, positrons and neutrinos are in thermal equilibrium in each sector separately, that is $T_{\nu} = T_e = T \; , \; T_{\nu}' = T_e' = T'$.\footnote{
For simplicity we write $T'_{\nu'} = T'_{\nu}$ , $T'_{e'} = T'_e$ , $T'_{\gamma'} = T'_{\gamma} = T'$ , $T_{\gamma} = T$.}
Using equations from (\ref{energy_density}) to (\ref{g_tot__def}) for particles in both sectors we are able to calculate the  DoF number in O or M worlds alone; but to work out the \it total \rm DoF number, that is summed on both worlds, we need the M temperature $T'$ as a function of the O one $T$ or viceversa.

As we neglect the entropy exchanges between the sectors (valid since there are only gravitational interactions between them), we can obtain both these functions using Def.~(\ref{mir-param}) and imposing $x=constant$ in:
\begin{eqnarray} \label{x_cons_1}
  x^3
 =\frac{s'_e + s'_{\gamma} + s'_{\nu}}{s_e + s_{\gamma} + s_{\nu}}
 =\frac{\left[ \frac{7}{8}q_e(T')+ q_{\gamma} + \frac{7}{8} q_{\nu} \right]T'^3} 
  {\left[  \frac{7}{8}q_e(T)+ q_{\gamma} + \frac{7}{8} q_{\nu} \right] T^3} \;,
\end{eqnarray}
where $q_{\nu} = 6$ and $q_{\gamma} = 2$, while $q_e (T)$ stands for
\begin{eqnarray} \label{}
  q_e (T) = \frac{8}{7} \; \frac{s_e(T)}{\frac{2\pi^2}{45} T^3} \;,
\end{eqnarray}
with $s_e(T)$ defined in (\ref{entropy_generic}), and we have used $T'_e = T'_{\nu} = T'_{\gamma} = T'$,  $T_e = T_{\nu} = T_{\gamma} = T$.
We have also used that, since particles and physics are the same in both sectors, the equilibrium distribution functions, and thus the spin-degeneracy factors, are the same:
$q_e = q_e'$, $q_{\gamma} = q_{\gamma}'$, $q_{\nu} = q_{\nu}'$.

Equation (\ref{x_cons_1}) can be numerically solved in order to obtain the function $T'(T)$ for every $T > T_{D\nu'}$.

\subsubsection{Phase~ $T_{D\nu} < T \leq T_{D\nu'}$}

\noindent At $T \simeq T_{D\nu'}$ M neutrinos decouple and soon after M electrons and positrons annihilate, transferring their entropy only to the M photons, and hence raising their temperature.
In the M sector the entropies of the system ($e'^\pm$,$\gamma'$) and of $\nu'$ are separately conserved.
Nevertheless, O photons and neutrinos still have the same temperature $T$.
Therefore we have two equations. The first one comes from the conservation of entropies in the M sector, so that their ratio is equal to the asymptotic value computed at high temperatures, when the $e'^+$-$e'^-$ annihilation process had not begun yet:
\begin{eqnarray} \label{m_dec}
  \frac{S_e'+S_{\gamma}'}{S_{\nu}'}
 =\frac{s_e'+s_{\gamma}'}{s_{\nu}'}
 &=&\frac{\frac{7}{8}q_{e}(T')+q_{\gamma}}{\frac{7}{8}q_{\nu}}
  \left(\frac{T'}{T_{\nu}'}\right)^3 \nonumber\\
 &=&{\frac{7}{8} \cdot 4 + 2 \over \frac{7}{8} \cdot 6}
 =\frac{22}{21} \;,
\end{eqnarray}
and its solution gives $T_{\nu}'$ as a function of $T'$. The second one is the conservation of ratio of entropies in two sectors:
\begin{eqnarray} \label{x_cons_2}
  x^3
 =\frac{\left[ \frac{7}{8}q_e(T')+ q_{\gamma} \right] T'^3 
  + \frac{7}{8} q_{\nu} T_{\nu}'\,^3} 
  {\left[  \frac{7}{8}q_e(T)+ q_{\gamma} + 
  \frac{7}{8} q_{\nu} \right] T^3} \;,
\end{eqnarray}
and its solution, obtained using Eq.~(\ref{m_dec}), gives $T'$ as a function of $T$.
Eq.~(\ref{x_cons_2}) is the same as Eq.~(\ref{x_cons_1}) but with $T_e' = T_{\gamma}' = T' \ne T_{\nu}'$.

\subsubsection{Phase~ $T \leq T_{D\nu}$}

\noindent At $T\simeq T_{D\nu}$ O neutrinos decouple and soon after O electrons and positrons annihilate. Now also in the O sector the entropies of the system ($e^\pm$,$\gamma$) and $\nu$ are separately conserved; therefore we need one more equation to work out the O neutrino temperature $T_{\nu}$ as a function of the O photon one $T$:
\begin{eqnarray} \label{}
  \frac{S_e+S_{\gamma}}{S_{\nu}}
 =\frac{\frac{7}{8}q_{e}(T)+q_{\gamma}}{\frac{7}{8}q_{\nu}}
  \left(\frac{T}{T_{\nu}}\right)^3 
 =\frac{22}{21} \;,
\end{eqnarray}
\begin{eqnarray} \label{}
  \frac{S_e'+S_{\gamma}'}{S_{\nu}'}
 =\frac{\frac{7}{8}q_{e}(T')+q_{\gamma}}{\frac{7}{8}q_{\nu}}
  \left(\frac{T'}{T_{\nu}'}\right)^3 
 =\frac{22}{21} \;,
\end{eqnarray}
\begin{eqnarray} \label{x_cons_3}
  x^3
 = \frac{\left[ \frac{7}{8}q_e(T')+ q_{\gamma} \right] T'^3 
 + \frac{7}{8} q_{\nu} T_{\nu}'\,^3}
 {\left[  \frac{7}{8}q_e(T)+ q_{\gamma} \right] T^3 
 + \frac{7}{8} q_{\nu} T_{\nu}^3} \;.
\end{eqnarray}
Eq.~(\ref{x_cons_3}) is the same as Eq.~(\ref{x_cons_2}) but with $T_e = T_{\gamma} = T \ne T_{\nu}$.

Once both O and M photon temperatures are known, it is straightforward to calculate the total energy and entropy densities; then, reversing equations (\ref{entropy}) and (\ref{energy}), we can work out the entropic $(q)$ and energetic $(g)$ number of DoF.
Calculations have been made for several different values of $x$, as reported in the following section.

%---------------------------------------------------------------------------------

\section{Numerical calculations}
\label{sec_Num_calc}

\begin{figure}
\includegraphics[scale=1.0]{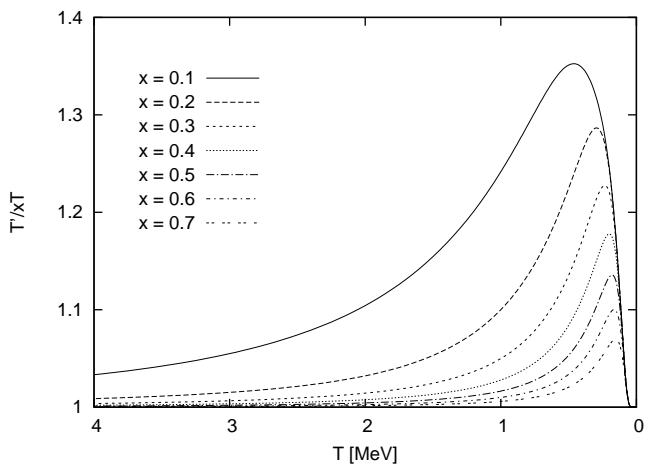}	
\caption{{\it The ratio $\frac{T'}{xT}$ for several values of $x$.
The asymptotic values of this ratio are $1$ for both high and low temperatures, as expected.}}
\label{fig_Tm_xT}
\end{figure}

The equations we introduced in the previous section have been numerically solved.
Since $x$ is a free parameter in our theory, several values have been used for 
it - from $0.1$ to $0.7$ with step $0.1$ or less.
In the extreme asymptotic cases $T \gg T_{D\nu'} \simeq \frac{T_{D\nu}}{x}$ or $T \ll T_{ann \: e^{\pm}}\simeq 1$ MeV we expect $q(T)\simeq q'[T'(T)]$; therefore, using Eq.~(\ref{xmir}), in this limits the ratio of M and O photon temperatures should be $x$, that is $\frac{T'}{xT} \simeq 1$.
Instead, when $T_{D\nu'} \gsim T \gsim T_{ann \: e^{\pm}}$ we expect $q'[T'(T)] \leq q(T)$ because the $e^+$-$e^-$ annihilation takes place before in the M world, leading to a decrease of $q'(T')$ and a corresponding increase of $T'$ in order to keep constant the entropy densities ratio; thus we expect $\frac{T'}{xT} > 1$.
%In conclusion, in this regime we should have $T'>xT$.
Later on, when the O electrons and positrons annihilate, they make even $T$ increase and thus the ratio $\frac{T'}{xT}$ decreases to the asymptotic value 
$1$.
These remarks have been numerically verified; the ratio $\frac{T'}{xT}$ is plotted in Fig.~\ref{fig_Tm_xT} for different values of $x$.

M $e^+$-$e^-$ annihilation happens at higher O temperatures for lower $x$ values, raising before the temperature of M photons; this results in a shift of the peaks of Fig.~\ref{fig_Tm_xT} towards higher T.
In addition, for lower $x$ the difference in T between the two annihilation processes is higher, so M photons have more time to raise their temperatures before O ones start to do the same, and this leads to the change of the shape of the curves.
%

%.................................................................................

\subsection{Number of degrees of freedom}
\label{sec_DoF_num}

Using equations from (\ref{energy_density}) to (\ref{entropy_generic}) to 
reverse (\ref{entropy}) and (\ref{energy}), we can work out the total number 
of O entropic ($\bar q$) and energetic ($\bar g$) DoF at any temperatures $T$.
We can apply the same procedure to work out the {\it standard}, that is in absence of the M sector, total $q_{std}$ and $g_{std}$, as well as the M ones $\bar q'$ and $\bar g'$, and the influence of a world on the other one.

At first approximation we expect $\bar q$ ($\bar g$) to have a cubic (quartic) dependence on $x$ when the temperature is not close to the $\nu$ decoupling and the $e^+$-$e^-$ annihilation phases: $\bar q = q_{std} (1+x^3)$,  $\bar q' = q_{std} (1+x^{-3})$, $\bar g = g_{std} (1+x^4)$, $\bar g' = g_{std} (1+x^{-4})$ (for an explanation see Ref.~\cite{Berezhiani:2000gw}).

In the following we present the results of accurate numerical calculations.
In Table~\ref{tab_DoF} some values are reported for special temperatures and several values of $x$.
As expected, the total DoF numbers are always higher than the standard and increase with $x$.
Moreover, the M sector values are higher than the O ones by a 
factor of order $x^{-3}$ (for $\bar q$) or $x^{-4}$ (for $\bar g$).

\begin{figure}
\includegraphics[scale=1.15]{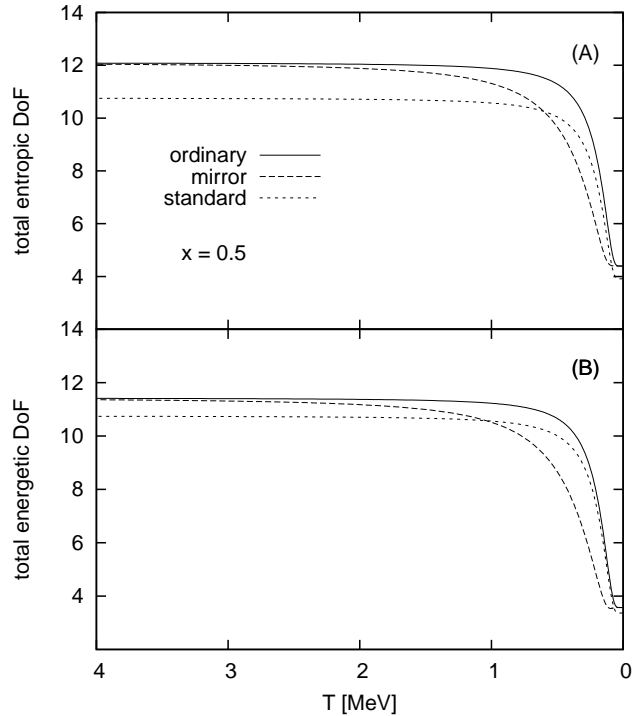}	
\caption{{\it Total entropic (panel A) and energetic (panel B) degrees of freedom computed in ordinary and mirror sectors and for the standard. The mirror values have been multiplied by $x^3$ ($q$) and $x^4$ ($g$) to make them comparable with the ordinary ones, since $\frac{T'}{xT} \sim 1$.}}
\label{fig_DoF_tot_x_05}
\end{figure}

In Fig.~\ref{fig_DoF_tot_x_05}A $q_{std}$, $\bar q$, $\bar q'$ are plotted for the intermediate value $x=0.5$; panel B of the same figure shows the corresponding values of $g$.
In the figure $\bar q'$ ($\bar g'$) has been scaled by a factor $x^3 $ ($x^4$); in this way the asymptotic values are the same than the O sector ones because at the extremes $\frac{T'}{T}=x$ (see Fig.~\ref{fig_Tm_xT}).
We can see that $\bar q$, $\bar g$, $\bar q'$ and $\bar g'$ have similar trends, but, due to $T'<T$, DoF in the M sector begin to decrease before (at higher $T$).

\begin{figure}
\includegraphics[scale=1.15]{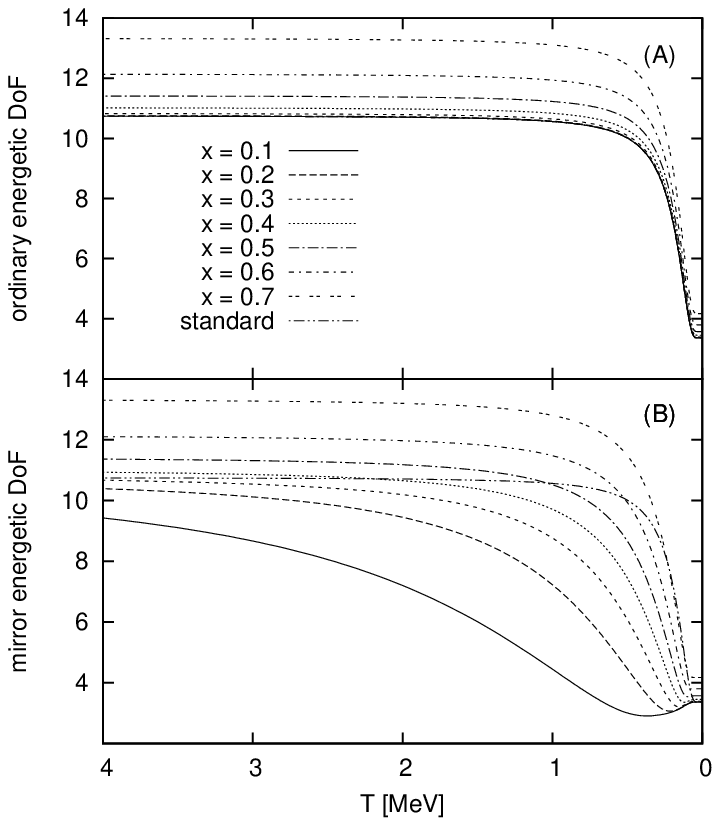}	
\caption{{\it Total energetic degrees of freedom in the mirror sector computed for several values of $x$ and compared with the standard.}}
\label{fig_ord_mir_DoF}
\end{figure}

In Fig.~\ref{fig_ord_mir_DoF} $\bar g$ in O and M sectors are plotted in comparison with the standard for several $x$ values (from $0.1$ to $0.7$ with step $0.1$).
The predicted quartic dependence of $g$ on $x$ at the extremes is proved correct.
In panel A we note that ordinary DoF with $x<0.3$ are practically identical to the standard case.
In addition, the plot shape does not change with $x$ in the O sector, while it does in the M one.
This sector evolves with temperature $T' \sim xT < T$; therefore, for lower $x$ the number of DoF starts before decreasing below the asymptotic value at high $T$.
The change of the shape of the plots in panel B is related to the same physical processes responsible of the analogous effect present in Fig.~\ref{fig_Tm_xT}, that is due to the $e^+$-$e^-$ annihilation in the M sector.
\begin{table}
\caption{Standard and non-standard total DoF numbers for several $x$ values in both O and M sectors. Temperatures are in MeV.}
\begin{ruledtabular}
\begin{tabular}{cccccc}
$T $ & $0.005$ & $0.1$ & $0.5$ & $1$ & $5$ \\
\hline
$q_{std}$ & 3.91 & 4.78 & 10.0 & 10.6 & 10.75 \\
$g_{std}$ & 3.36 & 4.30 & 10.0 & 10.6 & 10.75 \\
\hline
\multicolumn{6}{c}{$\mathbf{x = 0.1}$} \\
\hline
$T'$ & 0.0005 & 0.0107 & 0.0675 & 0.124 & 0.511 \\
$\bar q$ & 3.91 & 4.78 & 10.0 & 10.6 & 10.75 \\
$\bar q'$ & 3913 & 3913 & 4072 & 5522 & 10070 \\
$\bar g$ & 3.36 & 4.30 & 10.0 & 10.6 & 10.75 \\
$\bar g'$ & 33629 & 32894 & 30032 & 44430 & 98436 \\
\hline
\multicolumn{6}{c}{$\mathbf{x = 0.3}$} \\
\hline
$T'$ & 0.0015 & 0.0321 & 0.170 & 0.315 & 1.50 \\
$\bar q$ & 4.015 & 4.91 & 10.3 & 10.8 & 11.0 \\
$\bar q'$ & 149 & 149 & 261 & 347 & 406 \\
$\bar g$ & 3.39 & 4.34 & 10.1 & 10.6 & 10.8 \\
$\bar g'$ & 418.5 & 409 & 750 & 1082 & 1325 \\
\hline
\multicolumn{6}{c}{$\mathbf{x = 0.5}$} \\
\hline
$T'$ & 0.0025 & 0.0533 & 0.263 & 0.508 & 2.50 \\
$\bar q$ & 4.40 & 5.38 & 11.3 & 11.9 & 12.1 \\
$\bar q'$ & 35.2 & 35.5 & 77.3 & 90.5 & 96.5 \\
$\bar g$ & 3.57 & 4.58 & 10.7 & 11.2 & 11.4 \\
$\bar g'$ & 57.2 & 56.6 & 138 & 168 & 182 \\
\hline
\multicolumn{6}{c}{$\mathbf{x = 0.7}$} \\
\hline
$T'$ & 0.0035 & 0.0733 & 0.357 & 0.704 & 3.50 \\
$\bar q$ & 5.25 & 6.42 & 13.5 & 14.2 & 14.4 \\
$\bar q'$ & 15.3 & 16.3 & 36.9 & 40.6 & 42.0 \\
$\bar g$ & 4.17 & 5.35 & 12.4 & 13.1 & 13.3 \\
$\bar g'$ & 17.4 & 18.5 & 47.8 & 53.3 & 55.5 \\
\end{tabular}
\end{ruledtabular}
\label{tab_DoF}
\end{table}

%.................................................................................

\subsection{Number of neutrino families}
\label{sec_Nu_num}

\begin{figure}
\includegraphics[scale=1.0]{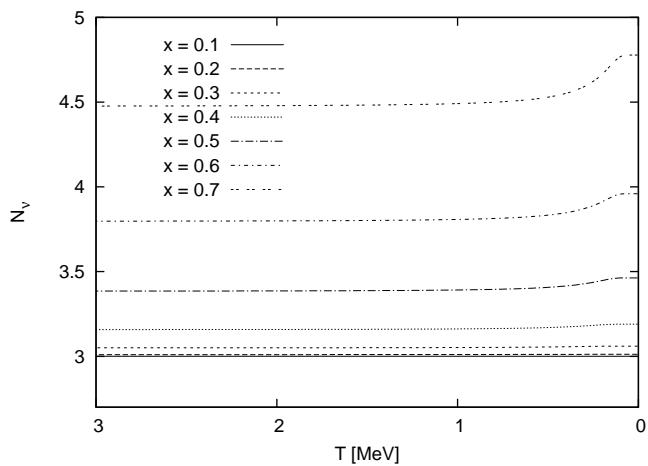}	
\caption{{\it Effective number of neutrino families $N_{\nu}$ in the ordinary sector for several $x$ values.}}
\label{fig_Nnu}
\end{figure}

We know that the SM contains three neutrino species; the possible existence of a fourth neutrino has been investigated for a long time, also using BBN constraints.
This is why in the literature one can often find bounds on the number of DoF in terms of {\sl effective} extra-neutrino number $\Delta N_{\nu} = N_{\nu} - 3$.
In general, the effective number of neutrinos $N_{\nu}$ is found assuming that all particles contributing to the Universe energy density, to the exclusion of electrons, positrons and photons, are neutrinos; in formula that means:
\begin{eqnarray} \label{Nnu_conversion}
  \bar g (T) = g_e(T) + g_{\gamma} +
    \frac{7}{8}\cdot 2 N_{\nu}\cdot\left(\frac{T_{\nu}}{T}\right)^4 \nonumber\\
      \Rightarrow
  N_{\nu} =
    \frac{\bar g (T) - g_e(T) - g_{\gamma}}{\frac{7}{8}\cdot 2}
    \cdot \left( \frac{T}{T_{\nu}} \right)^4 \;.
\end{eqnarray}
$N_{\nu}$ has been worked out using Eq.~(\ref{Nnu_conversion}) together with the results of previous numerical simulations; some data are reported in Table~\ref{tab_Nnu}, while plots for several $x$ values and any temperatures from $0$ to $3$ MeV are shown in Fig.~\ref{fig_Nnu}.
We stress that the standard values $N_{\nu} = 3$ is the same at any temperatures, while {\sl a distinctive feature of mirror scenario is that the number of neutrinos raises for decreasing temperatures}.
Anyway, this effect is not a problem; on the contrary it may be useful since recent cosmological data fits give indications for a number of neutrinos at recent times higher than at BBN epoch.
In Ref.~\cite{Mangano:2006ur} the authors found $N_\nu = 5.2^{+2.7}_{-2.2}$ using recent cosmic microwave background (CMB) and large scale structure (LSS) data.
We can use the aforementioned mirror feature together with the results of a previous work on (CMB) and (LSS) power spectra~\cite{Ciarcelluti:2003wm,Ciarcelluti:2004ip}, where author studied the dependence of the spectra on the parameters $x$ and $N_{\nu}$ for a flat Universe with mirror dark matter.
As can be easily evinced from Figures 11, 12, 14, 15 of Ref.~\cite{Ciarcelluti:2004ip}, an increase of the effective number of neutrinos is well mimicked by an increase of the parameter $x$, and the amounts of the respective increases are in accordance with what required to justify the data of CMB and LSS.
Thus, considering the sum of these effects, i.e. the raise in $N_{\nu}$ before and after BBN, and the similarity of CMB and LSS spectra of mirror models and standard $N_{\nu}$ with the ones obtained without mirror sector but with larger $N_{\nu}$, the mentioned discrepancy naturally disappears.
\begin{table}
\caption{Effective number of neutrinos in the ordinary sector for some special cases. Temperatures are in MeV.}
\begin{ruledtabular}
\begin{tabular}{dcccc}
$T$ & $x=0.1$ & $x=0.3$ & $x=0.5$ & $x=0.7$ \\
\hline
\multicolumn{5}{c}{{\bf ordinary sector}} \\
\hline
0.005 & 3.00074 & 3.05997 & 3.46270 & 4.77751 \\
0.1 & 3.00074 & 3.05997 & 3.46244 & 4.76829 \\
0.5 & 3.00074 & 3.05997 & 3.40706 & 4.52942 \\
1 & 3.00071 & 3.05202 & 3.39166 & 4.49133 \\
5 & 3.00063 & 3.04989 & 3.38430 & 4.47563 \\
\end{tabular}
\end{ruledtabular}
\label{tab_Nnu}
\end{table}

Similarly the effective number of neutrinos in the M sector can be worked out as:
\begin{eqnarray}
\label{Nnu_conversion_mir}
  N'_{\nu} =
    \frac{\bar g' - g'_e(T') - g'_{\gamma}}{\frac{7}{8}\cdot 2} 
    \cdot \left( \frac{T'}{T'_{\nu}} \right)^4 \;.
\end{eqnarray}
Once again these values are higher than the O ones, but now by a factor $x^{-4}$; they have been numerically computed and some special values are reported in Table~\ref{tab_Nnu_mir}. For lower $x$ this number can become very high, inducing relevant consequences on the primordial nucleosynthesis in the M sector, that is highly dependent on this parameter.
\begin{table}
\caption{Effective number of neutrinos in the mirror sector for some special cases. Temperatures are in MeV.}
\begin{ruledtabular}
\begin{tabular}{dcccc}
$T'$ & $x=0.1$ & $x=0.3$ & $x=0.5$ & $x=0.7$ \\
\hline
\multicolumn{5}{c}{{\bf mirror sector}} \\
\hline
0.005 & 74011 & 917.0 & 121.4 & 33.83 \\
0.1 & 62007 & 805.0 & 111.6 & 32.21 \\
0.5  & 61447 & 763.1 & 101.9 & 28.86 \\
1  & 61435 & 761.8 & 101.4 & 28.66 \\
5  & 61432 & 761.4 & 101.3 & 28.59 \\
\end{tabular}
\end{ruledtabular}
\label{tab_Nnu_mir}
\end{table}

\paragraph{Predictions for special temperatures}

It is possible to work out the asymptotic values of $N_{\nu}$ at $T \gg T_{D\nu}$ and $T \ll T_{ann \, e^{\pm}}$ in a simple way starting from the standard values:
\begin{eqnarray}
\label{g_extremes_std}
  g_{std}(T \gg T_{D\nu}) = 10.75\;;
\cr\cr
  g_{std}(T \ll T_{ann \, e^{\pm}}) \simeq 3.36 \;.
\end{eqnarray}
Without the M sector, we have, as expected:
\begin{eqnarray} \label{Nnu_Delta_Nnu_standard}
  N_{\nu} (T \gg T_{D\nu})  &=& \frac{10.75 - 2 - \frac{7}{8} \cdot 4}
    {{\frac{7}{8}\cdot 2}} = 3 \;;
\cr\cr
  N_{\nu} (T \ll T_{ann \, e^{\pm}}) &=& \frac{3.36 - 2}
    {{\frac{7}{8}\cdot 2}} 
    \cdot \left( \frac{11}{4} \right)^\frac{4}{3} = 3 \;.
\end{eqnarray}
While, when the M sector is present, we can use the previously mentioned quadratic approximation $\bar g = g_{std} (1+x^4)$ at the special temperatures we are considering; hence:
\begin{eqnarray} \label{Nnu_asynt}
  N_{\nu} (T \gg T_{D\nu})
    &=&\frac{10.75(1+x^4)-2-\frac{7}{8}\cdot 4}{{\frac{7}{8}\cdot 2}} \nonumber\\
    &=& 3 + 6.14~x^4 \;;
\cr\cr
  N_{\nu} (T \ll T_{ann \, e^{\pm}}) 
    &=&\frac{3.36 (1+x^4)- 2}{{\frac{7}{8}\cdot 2}} \cdot
       \left( \frac{11}{4} \right)^\frac{4}{3} \nonumber\\
    &\simeq& 3 + 7.40~x^4\;.
\end{eqnarray}
From the equations written above we can see that the rise $\Delta N_{\nu}$ is:
\begin{eqnarray} \label{N_nu_mir_approx}
  \Delta N_{\nu} 
    &=& N_{\nu} (T \ll T_{ann \, e^{\pm}}) - N_{\nu} (T \gg T_{D\nu}) \nonumber \\
    &=& x^4 \cdot \frac{1}{\frac{7}{8}\cdot 2} \left[ 3.36 
      \left( \frac{11}{4} \right)^\frac{4}{3} - 10.75 \right] \nonumber \\
    &\simeq& 1.255 \cdot x^4 \;.
\end{eqnarray}
This leads to $N_{\nu}$ higher than the standard in presence of the M world and to a further rise of this parameter at low temperatures.
If we assume a conservative limit on the effective number of neutrino families $N_{\nu} (T \gg T_{D\nu}) \le 4$, this implies $x \le 0.635$ and $N_{\nu} (T \ll T_{ann \, e^{\pm}}) \le 4.26$.
For some interesting values of the parameter $x$ we obtain:
\begin{eqnarray} \label{}
  x = 0.7 \hspace{0.5cm} \Rightarrow \hspace{0.5cm}
    N_{\nu}(T \gg T_{D\nu}) \simeq 4.5
    \hspace{0.5cm} &\mathrm{and}& \hspace{0.5cm} \nonumber\\
    N_{\nu} (T \ll T_{ann \, e^{\pm}}) \simeq 4.8 \;;
\cr\cr
  x = 0.6 \hspace{0.5cm} \Rightarrow \hspace{0.5cm}
    N_{\nu} (T \gg T_{D\nu}) \simeq 3.8
    \hspace{0.5cm} &\mathrm{and}& \hspace{0.5cm} \nonumber\\
    N_{\nu} (T \ll T_{ann \, e^{\pm}}) \simeq 3.96 \;;
\cr\cr
  x = 0.3 \hspace{0.5cm} \Rightarrow \hspace{0.5cm}
    N_{\nu} (T \gg T_{D\nu}) \simeq 3.05
    \hspace{0.33cm} &\mathrm{and}& \hspace{0.5cm} \nonumber\\
    N_{\nu} (T \ll T_{ann \, e^{\pm}}) \simeq 3.06 \;.
\end{eqnarray}
These rough estimates are in good agreement with the asymptotic numerical values computed before and after BBN; nevertheless, the previous detailed analysis of the evolution of this quantity is crucial for future studies of the nucleosynthesis in both sectors.

%---------------------------------------------------------------------------------

\section{Conclusions}
\label{sec_conclusions}

In this paper we have investigated the consequences of the existence of a mirror sector of particles and interactions with exact parity symmetry on the thermodynamics of the early Universe for temperatures below $\sim 100$ MeV.
In particular we studied in detail the evolution of the entropic and energetic degrees of freedom in both sectors.

To perform these calculations we first found the system of equations governing the thermodynamical equilibria in the two sectors, and then we numerically solved them in order to obtain the mirror temperature $T'$ corresponding to any ordinary temperature $T$; once these temperatures were known, we were able to work out the entropic and energetic densities, and thus the numbers of degrees of freedom and the effective numbers of neutrino families in both sectors.
We found that the evolution of the ordinary sector is influenced by the mirror particles, that provide extra degrees of freedom, and, due to the electron-positron annihilation process happening in the mirror sector, changes with time for temperatures around the BBN epoch.
This makes crucial the present investigation in order to study in details the BBN process in presence of mirror dark matter.
Furthermore, the special feature that models containing the mirror sector change the number of equivalent neutrinos during the Universe evolution, if considered together with preexisting models of cosmic microwave background (CMB) and large scale structure power spectra with mirror dark matter, furnishes a possible interpretation of the discrepancy in the effective number of neutrino families indirectly ``observed" at BBN and CMB epochs.

In the mirror sector we obtained very high values of degrees of freedom, as predicted, and again they are variable for temperatures below some MeV. These data provide the necessary input for a detailed study of BBN in the mirror sector.
It is required in order to obtain the exact amounts of primordial heavy mirror elements, that are necessary for a correct interpretation of non-linear processes of structure formation and of the new DAMA/LIBRA results.

%---------------------------------------------------------------------------------

\begin{acknowledgments}

We thank Zurab Berezhiani for interesting discussions and comments. 
P. C. acknowledges support from the Belgian fund for scientific research (FNRS).

\end{acknowledgments}

%---------------------------------------------------------------------------------

\newpage %Just because of unusual number of tables stacked at end
\bibliography{g_mir_aps}% Produces the bibliography via BibTeX.

%---------------------------------------------------------------------------------

\end{document}